\documentclass[twocolumn,showpacs,preprintnumbers,amsmath,amssymb]{revtex4}
\usepackage{graphicx}
\usepackage{bm}

\begin{document}

\title{ The best sequence for Parrondo games}

\author{Luis Dinis}
\affiliation{Grupo Interdisciplinar de Sistemas Complejos (GISC) and Departamento de F\'{i}sica At\'omica, Molecular y Nuclear.\\ Universidad Complutense de Madrid. Ciudad Universitaria E-28040 Madrid (Spain)}
\email{ ldinis@fis.ucm.es}
\date{\today}
\begin{abstract}
An algorithm based on backward induction is devised in order to compute the optimal sequence of games to be played in Parrondo games. The algorithm can be used to find the optimal sequence for any finite number of turns or in the steady state, showing that ABABB\ldots is the sequence with the highest steady state average gain. The algorithm can also be generalised to find the optimal adaptive strategy in a multi-player version of the games, where a finite number of players may choose, at every turn, the game the whole ensemble should play. 
\end{abstract}

\maketitle

\section{Introduction}
Rectification of thermal fluctuations has become a major topic in non equilibrium Statistical Physics. Ajdari and Prost \cite{ajdari} discovered in 1993 a Brownian ratchet mechanism, afterwards named by Astumian and Bier the {\em flashing ratchet} \cite{astumian}. In 1996, Parrondo \cite{parrseminal} showed that this rectification mechanism also works when spatial and time degrees of freedom of the brownian particle are discrete, as in the chance games thereafter known as Parrondo games\cite{abbotharmer}:
 two separate losing games that can be combined following a random or periodic strategy resulting in a winning game.

The games have received attention in several disciplines, ranging from  quantum game theory \cite{quabbott,quantum}, non-linear or chaotic dynamics \cite{kocarev,arena}, economics \cite{boman,jbf} and biology \cite{jtb,davies}.
However, the question on how to combine the games  to get the highest increase in capital, was still open.
Sequences up to period 12 have been studied using symbolic manipulators \cite{vellerman}, and the periodic sequence ABABB (or any of its permutations) has come up as the best in the sense that it provides the highest returns in the stationary state. In this paper we show that ABABB\ldots is indeed the best sequence by applying Bellman's optimality criterion \cite{bellman} and backward induction. 

%


Recently, various multi-player versions of the games have been proposed, giving rise to counter-intuitive phenomena resembling those observed in game, control and optimization theories or economics. For instance, it turns out that greedy algorithms 
or strategies may lead to suboptimal or even losing solutions \cite{dinisepl,dinisphysA,sotillo}.

However, it is worth noting that, contrary to what happens in many of the models used in economy or game theory, the behavior of Parrondo games is of a purely stochastic nature, therefore making them a good system to help understand the role of fluctuations and optimization in those systems where stochastic dynamics is relevant. As an example, it has also been shown that a related phenomenon may occur in a feedback controlled collective flashing ratchet \cite{cao,controlepl}.
Also in this context, the problem of finding an optimal protocol or strategy in a system where fluctuations have a major role has received attention lately in the field of finite-time thermodynamics and fluctuation theorems \cite{Seifert}.

Finally, one can think of the problem of finding the best sequence in an alternative way. Imagine an infinite number of independent players who play against a casino with the only restriction that all of them must play the same game (A or B) at every turn $t$. That is, the decision to play A or B at $t$ is taken collectively. If some information about the state of the system is known at $t$ (we will later see which is the information needed), an optimal way of choosing A or B can be found so that the average capital is the maximum possible. This is an {\em adaptive strategy} in the sense that the choice taken can be adapted to the current state of the system; the interested reader may find a formal proof that backward induction can be used to calculate the best adaptive strategy explicitly in \cite{behrends}.
 It then turns out that this optimal adaptive strategy makes the players use the sequence ABABB$\ldots$ in the long run. The average capital of an infinite ensemble of independent players is related, due to the law of large numbers, to the average of one player playing the same exact sequence of games. Hence, this will allow us to state that ABABB$\ldots$ is the best periodic sequence for the Parrondo Games.

We thus provide an example that an open-loop control problem (a control problem without information about the system) can be solved as a closed-loop optimization problem over an infinite collection of identical systems in which the information about their state may be used, an example that may be relevant for stochastic control theory.  

The paper is organised as follows: we begin by briefly reviewing game rules and evolution equation in section~\ref{sec:juegos}. In section~\ref{sec:problema} we state the problem and in
 sections~\ref{sec:solucion1} to \ref{sec:resultados},  we show how to find the best possible sequence to play in the long run for the original Parrondo games. 
In fact, the solution is  more general, and consists in finding the best sequence of games for any finite number of turns. 
The algorithm can be easily generalised to find the best way of choosing games for an arbitrary number of players (and number of turns to play), i.e. the best adaptive strategy for any number of players and any number of turns; this is done in section~\ref{sec:finite}. Section~\ref{sec:ppp} is devoted to the application of the algorithm to the Primary Parrondo Paradox. Concluding remarks can be found in section~\ref{sec:conclus}.     

\section{The games\label{sec:juegos}.}
Parrondo games can be stated as two simple coin tossing games, A and B. 
Game A is played with a coin slightly biased so that the probability to win  is less than one half, that is $p_A=1/2-\epsilon$, with $\epsilon$ a small positive number. Let $X(t)$ be the capital of the player in turn $t$. The average capital $\langle X(t)\rangle$ evolves with the number of turns as
\begin{equation}
\langle X(t+1)\rangle=\langle X(t)\rangle+2p_A-1,
\end{equation} 
 and therefore $\langle X(t)\rangle$ decreases with the number of turns. In this sense, we will call the game A a {\em losing} game. Anagolously, a {\em winning} game will be one in which $\langle X(t)\rangle$ increases with $t$.

Game B is played with two biased coins, say, the ``good'' and the ``bad'' coins. If the capital of the player is a multiple of 3, she must play the bad coin, which has a probability of winning $p_b=1/10-\epsilon$. Otherwise she tosses the good coin and wins with probability $p_g=3/4-\epsilon$. 
It can be shown \cite{cisneros} that these rules make B also a losing game in the long run. The paradox arises when alternating game A and B either in a random or periodic fashion, as this yields an average capital that increases with $t$, provided $\epsilon$ is small enough.  

If game B is played at turn $t$, the capital of the player changes as
\begin{equation}
\langle X(t+1)\rangle=\langle X(t)\rangle+2p_\text{winB}(t)-1,
\end{equation}
where $p_\text{winB}$ is the probability to win in game B, which depends on the capital of the player in turn $t$. 
More precisely, it only depends on the probability $\pi_0(t)$ that the player has a capital multiple of three in the $t$'th turn. With this definition, game B rules imply
\begin{equation}
p_\text{winB}(t)=\pi_0(t)p_b+(1-\pi_0(t))p_g
\end{equation}
To compute $\pi_0(t)$, one can define $\pi_1(t)$ and $\pi_2(t)$ as the probabilities that the capital is a multiple of 3 plus 1 or plus 2 respectively, and $\bm{\pi}(t)\equiv(\pi_0(t),\pi_1(t),\pi_2(t))^t$. Alternatively, $\langle X(t)\rangle$ can be interpreted as the average over the population of an infinite ensemble of independent players and $\pi(t)$ as fraction of players instead of probabilities.  In either case, the following evolution equation applies
\begin{eqnarray}
\bm{\pi}(t+1)& =&\Pi_{B}\bm{\pi}(t), \nonumber 
  \text{ with } \\ \Pi_B&=&\left(
\begin{array}{ccc}
0 & 1-p_g & p_g\\
p_b & 0 & 1-p_g\\
1-p_b & p_g & 0
\end{array}
\right)
.  
\label{eq:evolA}
\end{eqnarray}
Finally, game A can be expressed in the same terms and its evolution equation is
\begin{eqnarray}
\bm{\pi}(t+1)&=&\Pi_{A}\bm{\pi}(t), \text{ where }\nonumber\\
 \Pi_A&=&\left(
\begin{array}{ccc}
0 & 1-p_A & p_A\\
p_A & 0 & 1-p_A\\
1-p_A & p_A & 0
\end{array}
\right)
.
\label{eq:evolB}
\end{eqnarray}


Due to normalisation of probabilities, $\pi_0+\pi_1+\pi_2=1$, so the system state is fully determined by $(\pi_0,\pi_1)$. Hence, the region accessible to the system can be represented as a rectangular triangle of side 1 in $(\pi_0,\pi_1)$ space. 
\section{The problem\label{sec:problema}}

The expected gain $g(\pi(t))$ when playing game A or B in turn $t$ is defined as
\begin{equation}
g(\pi_0(t))\equiv \langle X(t+1)\rangle - \langle X(t)\rangle,
\end{equation}
and may have two different expressions
\begin{equation}
g(\pi_0(t))=
\begin{cases}
g^A\text{ if A is played at } t\\
g^B\text{ if B is played at }t
\end{cases}
\end{equation}
with 
\begin{eqnarray}
g^A&\equiv& 2p_A-1\\
g^B(\pi_0(t))&\equiv& 2[\pi_0(t)p_b+(1-\pi_0(t))p_g]-1
\end{eqnarray}
as shown in the previous section. The total gain after $T$ turns of the games is
\begin{equation}
G_T=\sum_{t=1}^Tg(t)
\end{equation}
and it is the target function we aim to maximise.

Let $\alpha_t$ be a parameter that may have values A or B to mark the game to be played at turn $t$, and $(\alpha_1,\alpha_2,\ldots,\alpha_n)$ a sequence of decisions to play A or B.
The problem we are faced with can be stated in the following way:

``Find the sequence of decisions $(\alpha_1,\alpha_2,\ldots,\alpha_n)$ so that $G_T$ attains its maximum value, with the restriction that $\bm{\pi}(t)$ evolves as
\begin{equation}
\bm{\pi}(t+1)=\Pi_A\bm{\pi}(t), \text { if }\alpha_t=\text{A}
\label{eq:evol1}
\end{equation}
or
\begin{equation}
\bm{\pi}(t+1)=\Pi_B\bm{\pi}(t), \text { if }\alpha_t=\text{B},
\label{eq:evol2}
\end{equation}
provided $\bm{\pi}(t)$ is known".

\section{Formal solution\label{sec:solucion1}}

Since any decision $\alpha_t$ affects the next, the best way to approach this problem is by proceeding backwards. The last decision will not affect any other, so it makes sense to start with that one. Once we know how we should proceed when we arrive to the last step, we can use that information to try and find out what is the best thing to do in the last but one turn of the games and so on.

Let us call $\hat G_n(\bm{\pi})$ the maximum possible value of the expected gain when there still are $n$ turns left \footnote{We will later see that this only depends on the number of turns left and the state of the system $\bm{\pi}$ in that moment}. At this stage, we could choose to play game A or B. If we do the former, the state of the system changes to $\Pi_A\bm{\pi}$ and the gain obtained in that step is $g^A$. The highest expected gain attainable by choosing A is then
\begin{equation}
g^A+\hat G_{n-1}(\Pi_A\bm{\pi}), 
\end{equation}
because $\hat G_{n-1}(\Pi_A\bm{\pi})$ is by definition the maximum gain that can be attained provided there are $n-1$ turns left and the system is in state $\Pi_A\bm{\pi}$. This can be stated more formally using Bellman's optimality criterion which assures that ``an optimal sequence has the property that, whatever the initial state and decision may be, the remaining decisions constitute and optimal sequence with respect to the state resulting from the first decision'' \cite{bellman}. 

If on the other hand we choose to play B when there are $n$ turns left, the maximum expected gain is 
\begin{equation}
g^B(\pi_0)+\hat G_{n-1}(\Pi_B\bm{\pi}).
\end{equation}
Those are the only two possibilities and therefore
\begin{eqnarray}
\label{eq:recursion}
&&\hat G_n(\bm{\pi})=\\&=&\max\left\{g^A+\hat G_{n-1}(\Pi_A\bm{\pi}),\ g^B(\pi_0)+\hat G_{n-1}(\Pi_B\bm{\pi})\right\}. \nonumber
\end{eqnarray}
It is now clear that information about the state $\bm{\pi}$ of the system is needed in order to maximize the gain, as stated in the introduction.

Given the state $\bm{\pi}$, the optimal decision at turn $t=T-n+1$ ($n$ to the end) is to play A if the maximum corresponds to the first term in expression \eqref{eq:recursion} and B otherwise, so the optimal decision $\alpha_{T-n+1}$ is A in the first case and B in the second. 
Thus, each point in the state space can be related either to game A or B, creating a map that can be used at turn $T-n+1$ to decide which game to play in order to have the highest expected gain. In fact, due to linearity of the expressions involved, the state space is always divided in two connected regions, one for game A and the other for game B (see Fig.~\ref{fig:4pasos} for some examples).

Finally, when there is just one turn left, the best choice is to play game A if $g^A>g^B(\pi_0)$ and B otherwise \footnote{One could also choose to play game $g^A\geq g^B(\pi_0)$. This will affect which we consider the optimal sequence among various with identical average gain in some special cases for a finite number of turns and initial conditions. It will nevertheless yield the same optimal steady state sequence since the points in the limit cycle of ABABB\ldots lay far away from the boundary of the regions A and B}. Hence,
\begin{equation}
\hat G_1(\bm{\pi})=\max\left\{g^A,g^B(\pi_0)\right\},
\label{eq:paso1}
\end{equation}
which completes the induction algorithm.


\section{Numerical solution\label{sec:solucion2}}
\begin{figure*}[htbp]
\includegraphics[width=0.8\textwidth]{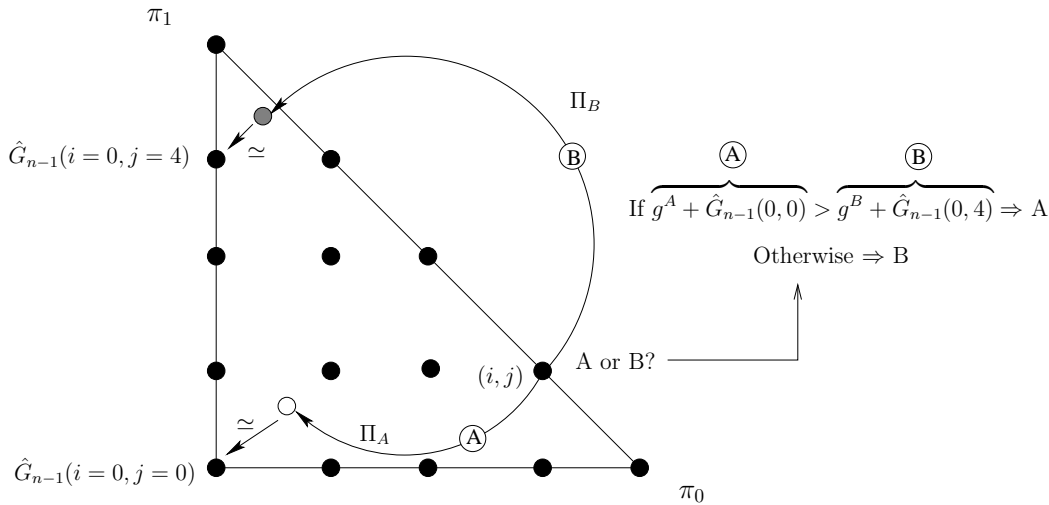}
\caption{\label{fig:gridninfinitoR} Evaluation of $\hat G_n(\pi_0^i,\pi_1^j)$ and computation of the optimal choice of game for state $(\pi_0^i,\pi_1^j)$ with $n$ turns left. }
\end{figure*}

The procedure explained in the preceding section can be readily turned into a recursive numerical algorithm to compute the maximum expected gain, given an starting state of the system $\bm{\pi}(1)$. 
Though the algorithm is simple to program, it requires as many operations as a brute force approach (that is, systematic evaluation of the $2^T$ possible sequences) due to recursion,  and what is even worse, does not provide any information about the solution for any other initial condition $\bm{\pi}(1)$. To tackle this, we used a different approach:
\begin{enumerate}
\item Define a grid in space $(\pi_0,\pi_1)$ as shown in Fig.~\ref{fig:gridninfinitoR}. Denote each point as $(\pi_0^i,\pi_1^j)$, or simply $(i,j)$, with $i,j\in\mathcal{Z} $. 
\item Set $n=1$. Evaluate $\hat G_1(i,j)$ in every point of the grid. 
\item Through evaluation of $\hat G_1(i,j)$, a map associating each point of the grid with the optimal choice (A or B) can be created.
\item \label{it:evalgnmenos1} Increase $n$ in one unit. Evaluate $\hat G_n$ in every point of the grid. To do that, values of $\hat G_{n-1}$ in points $\Pi_A (\pi_0^i,\pi_1^j,1-\pi_0^i-\pi_1^j)^t$ and $\Pi_B (\pi_0^i,\pi_1^j,1-\pi_0^i-\pi_1^j)^t$ which fall outside the grid would be needed. Approximate $\hat G_{n-1}$ in each of that points for the value in the closest point down-left in the grid, whose value is already known.
\item \label{it:evalgn} Evaluation of $\hat G_{n}$ provides a map associating each point of the grid to the optimal choice of the game in that point, when there are $n$ turns remaining.
\item Repeat steps \ref{it:evalgnmenos1} to \ref{it:evalgn} until $n=T$, the total number of turns.
\end{enumerate}

In step \ref{it:evalgnmenos1}, $\hat G_n$ is evaluated in every point of the grid for all $n$, although there is in principle no need to do so since the mappings defined by evolution Eqs.~\eqref{eq:evolA} and \eqref{eq:evolB} are contractive~\cite{behrends} and not every point can be reached by evolution after game A or B are played. However, by computing $\hat G_n$ at every point, we will obtain the solution not only for the optimal sequence of length $T$ but also of lengths $T-1, T-2,\ldots$ in just one run of the algorithm. Moreover, all of these solutions are valid for any initial condition.

Steps \ref{it:evalgnmenos1} and \ref{it:evalgn} are represented schematically in Fig.~\ref{fig:gridninfinitoR}. It is worth noting that the use of the approximation in step \ref{it:evalgnmenos1} implies the algorithm uses a time proportional to $T$ to compute the solution.

\section{Results\label{sec:resultados}. From the maps to the sequence.}
%

\begin{figure}[tbp]
\includegraphics[width=0.4\textwidth]{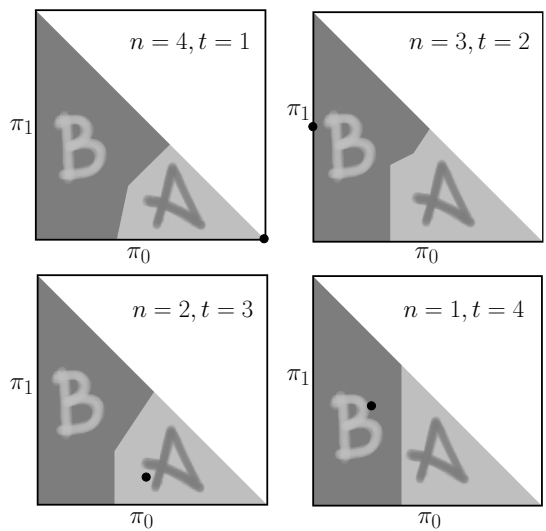}
\caption{\label{fig:4pasos} Maps and evolution for $T=4$, for turns $t=1,2,3,4$. $n=T-t+1$ denotes the number of turns left. The black circles mark the evolution of state $(\pi_0,\pi_1)$ when playing ABAB, starting from initial condition $(1,0)$.}
\end{figure}

As an example, imagine one player who is going to play the games four times ($T=4$) and whose initial capital is a multiple of 3 ($\pi_0=1, \pi_1=0$). Fig~\ref{fig:4pasos} shows the maps calculated using the aforementioned algorithm with a grid of $2000\times 2000$ points and $\epsilon=0$. For the remaining part of the article, I will take $\epsilon=0$ for simplicity. Choosing a small $\epsilon$ may shift the boundaries of the maps slightly. What is the sequence with the highest average gain provided the initial condition is $\pi_0=1$?

The map corresponding to $n=4$, indicates that for $(\pi_0,\pi_1)=(1,0)$ the best choice is game A. Game A takes $(\pi_0,\pi_1)$ from its initial value to $\Pi_A(1,0,0)^t=(0,0.5,0.5)^t$. Now there are 3 turns left and map $n=3$ shows that game B should be chosen, because $(0,0.5)$ falls in the region B, as shown in Fig.~\ref{fig:4pasos}. Game B takes the system to $(0.5,0.125)$ in region A of map $n=2$, and finally A takes it to $(0.25,0.4375)$, in region B of the last map. Therefore, sequence ABAB is the one with the highest expected gain.
Proceeding in this manner one can compute the optimal sequence for any number of turns $T$ and any initial condition. 

For a finite number of turns, some sprinting behavior can be observed that resembles the sprint effect at the beginning and the end of the time interval in general optimization problems \cite{Seifert,dinisepl}. The optimal sequence usually consists of several repetitions of the ABABB motif flanked by brief pieces of other sequences, as for example optimal sequence for 21 turns with initial condition (1,0): AB ABABBABABBABABB ABBABB (the spaces have been added to help identify the different parts).

In order to find the sequence with the highest {\em stationary} average gain, the behavior of the maps for $n\gg 1$ must be analysed. 

\begin{figure*}[htbp]
\includegraphics[width=0.8\textwidth]{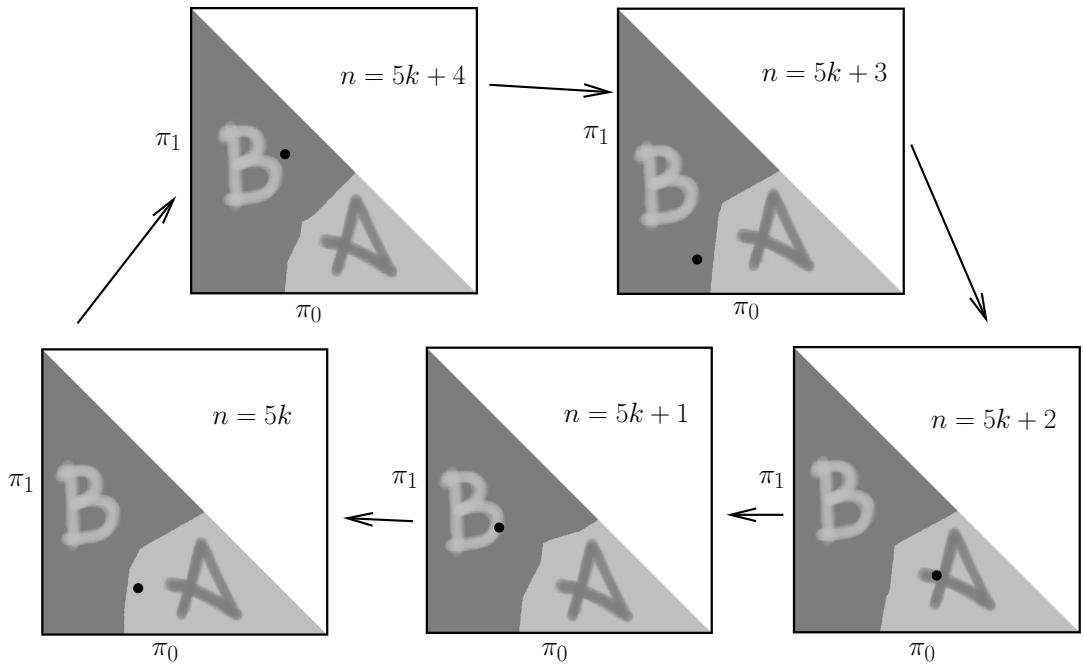}
\caption{\label{fig:5mapas} Stationary state maps for the optimal gain and their corresponding limit cycle. These 5 maps repeat periodically for $n\gg1$. To attain steady state, we set $k\gg 1(\Rightarrow n\gg 1)$ (still many turns left) and $t\gg1$ (many turns already played). To obtain the optimal sequence in the stationary state, the prescription indicated by the maps should be followed, using each map in the order indicated by the arrows. 
}
\end{figure*}

A regular pattern soon appears when increasing $n$, and the maps describing the regions in which to play A or B do not become completely independent of $n$ but change periodically with $n$, converging to a cycle of the 5 different maps depicted in Fig.~\ref{fig:5mapas}.
This means that 
one should follow the prescription contained in these maps, in a cyclic way and in order of decreasing $n$, as indicated by the arrows in Fig.~\ref{fig:5mapas}.

After a number of runs and irrespective of the initial condition, a steady state is attained where the player ends up playing the sequence ABABB\ldots, and $(\pi_0,\pi_1,\pi_2)$ follows the stationary five point cycle of this periodic sequence. The cycle is also shown in Fig.~\ref{fig:5mapas}. Hence, sequence ABABB\ldots is the one with the highest average gain in the long run.
\section{Finite number of players\label{sec:finite}}
In this section I turn to a different but related problem. Imagine now there is a single player which is allowed to decide whether to play A or B {\em depending on her actual state 0, 1 or 2}, as opposed  to our previous problem in which the player chose the sequence of games beforehand and had to keep to it irrespectively of the outcome of the games. Which is the way of choosing the games so that the average gain is the highest? The answer is quite trivial, the optimal choice can be expressed as ``every time you play, choose A if your state is zero, B otherwise''. As stated in the Introduction, this kind of recipe to choose the game depending of the state is usually termed a {\em strategy}. By using the former strategy, the player completely avoids playing with the bad coin and her average gain is maximal. 

The problem becomes more interesting if a finite ensemble of $N$ players has to play independently against a casino with the constraint that all of them must play the same game. Since the players will be in different states in general, either game will be a good choice only for some part of them. Which is then the optimal strategy? To answer this question we can use the previous algorithm, with a slight modification.
 
For a finite number of players, the state of the whole system is given by the fraction of players in any of the 3 possible states, that is, $\bm{\pi}(t)=(N_0,N_1,N_2)/N$, where $N_i$ is the number of players with a capital multiple of 3 plus $i$, and $N$ the total number of them. The main difference  is that now evolution of $\bm{\pi}(t)$ is stochastic, as opposed to the deterministic evolution described in section~\ref{sec:juegos} \footnote{If the ensemble were composed of an infinite number of players, the evolution could be described by the same Eqs.~\eqref{eq:evolA} and \eqref{eq:evolB} and the system would behave exactly as a single player since all the players must choose the same game. The optimal strategy for an infinite ensemble would coincide then with the maps obtained in section \ref{sec:resultados}}. Due to normalization, the state is sufficiently determined by $(N_0(t),N_1(t))$.

The expected gains for one turn $g^A$ and $g^B$, and also $\hat G_1(\pi_0)$, are defined as in the previous sections. However, to compute $\hat G_n(N_0/N,N_1/N)$, we must take into account the stochastic evolution of $(N_0,N_1)$. For example, if we start with a distribution of capitals $(N_0,N_1)=(i,j)$ and our first choice is A, we will in average obtain a gain

\begin{equation}
g^A+\sum_{(k,l)} p^A_{(i,j)\rightarrow(k,l)}\hat G_{n-1}(k/N,l/N),
\end{equation}
where $\hat G_{n-1} $ is weighted with the probability $p^A_{(i,j)\rightarrow(k,l)}$ to jump from state $(i,j)$ to $(k,l)$ in game A. An analogous expression can be written also for game B. Once the transition probabilities are computed using the rules of the games, the rest of the algorithm can be applied as described in section~\ref{sec:solucion2}.

The results obtained for a finite number of players agree with the currently available analytical solutions for the steady state (up to $N=3$ in Ref.~\cite{tesisbart}). Furthermore, this algorithm  can be used to compute the solution for more than a hundred players in a PC. Finally, it is worth mentioning that the algorithm allowed us to state that it is impossible to devise a strategy that gives the optimal average gain in the steady state irrespective of the number of players. The computation of the optimal ones for 25 and 100 players show that they differ in the game to be chosen in some points.

\section{Application to Primary Parrondo Paradox\label{sec:ppp}}

The algorithm can be also successfully applied to the Primary Parrondo Paradox (PPP) both to obtain the best a priori sequence for a single player or the optimal strategy in the multi-player PPP version with ``the freedom of choosing the common next game for the players'' \cite{ppp}.

PPP consists also in two games, A and B, with probabilities which depend in general on the capital of the player. Two possible states are defined, capital odd or even, and there is a probability to remain with the same capital after playing either of the games. The probability to win, stay with same capital and lose are given in table \ref{tab:probs} for both games. The state of the system is sufficiently defined giving $P_1$, the probability that the players has an even capital. With these probabilities one can show that $P_1=1/3$ always after playing B, and $P_1=1/2$ after playing A, irrespective of the previous state, a property known as superstability.

\begin{table}[b]
\begin{tabular}{|c|c|c|c||c|c|c|c|}\hline
{\bf A} & win & stay & loose & {\bf B} & win  & stay  & loose \\ \hline
odd & 1/4 & 1/2 & 1/4 & odd & 1/9 & 2/3 & 2/9\\ \hline
even & 1/4 & 1/2 & 1/4 & even & 4/9 &1/3 &2/9 \\\hline
\end{tabular}
\caption{\label{tab:probs} Probabilities to win, stay with the same capital or lose for games A and B in the Primary Parrondo Paradox.}
\end{table}

Due to the superstability of the PPP, which greatly simplifies the behavior of the system, the optimal sequence can be found analytically by applying Bellman's optimality criterion. 

The average gains in one turn are  $g_A=0$ and $g_B=1/3P_1-1/9$ for game A and Brespectively.
One can easily show that 
\begin{equation}
\hat G_{1}(P_1)=
\begin{cases}
0 \text{ if } P_1\leq \frac{1}{3} \text{ (play A)}\\
\frac{1}{3}P_1-\frac{1}{9}\text{ if } P_1>\frac{1}{3} \text{ (play B)}
\end{cases}
\end{equation}

\begin{equation}
\hat G_{2}(P_1)=
\begin{cases}
0 \text{ if } P_1< \frac{1}{2} \text{ (play A)}\\
\frac{1}{3}P_1-\frac{1}{9}\text{ if } P_1\geq\frac{1}{2} \text{ (play B)}.
\end{cases}
\end{equation}

Then, $\hat G_{3}$ can be computed from $\hat G_{2}$ yielding
\begin{equation}
\hat G_{3}(P_1)=
\begin{cases}
\frac{1}{18} \text{ if } P_1\leq \frac{1}{3} \text{ (A)}\\
\frac{1}{3}P_1-\frac{1}{9}+\frac{1}{18} \text{ if } P_1>\frac{1}{3} \text{ (B)}.
\end{cases}
\end{equation}
Therefore, $\hat G_3=\hat G_1+\frac{1}{18}$ and $\hat G_4$ necessarily has the same structure as $\hat G_2$, prescribing to also to choose A if $P_1<1/2$ and B otherwise; the computation of $\hat G_5$ yields the same prescription as $\hat G_3$, etc. Consequently, the optimal choice will only depend on whether $n$ is odd or even and can be expressed as:

\begin{equation}
n \text{ odd}\Rightarrow
\begin{cases}
\text{play A if } P_1\leq\frac{1}{3}\\ 
\text{play B if } P_1>\frac{1}{3} 
\end{cases}
\end{equation}

\begin{equation}
n \text{ even}\Rightarrow 
\begin{cases}
\text{play A if } P_1<\frac{1}{2}\\
\text{play B if } P_1\geq\frac{1}{2}
\end{cases}
\end{equation}
These prescriptions together with the evolution that takes $P_1$ to 1/3 when B is played and to 1/2 in case A is played, yield the sequence ABAB\ldots as the optimal sequence, in agreement with \cite{ppp}.

Regarding the multi-player version of the PPP with strategy, the numerical algorithm was modified to take into account the different transition probabilities of the PPP and the fact that there are only 2 possible states instead of 3. The optimal strategies provided by the modified algorithm have been checked proving identical to those reported in Ref.~\cite{optippp}, that is up to $N=10$. Moreover, using the algorithm I was able to compute the optimal strategy up to $N=100$.

\section{Conclusions\label{sec:conclus}}
Backward induction allowed us to compute the best sequence of games for any number of turns $T$ in time proportional to $T$. The algorithm shows ABABB\ldots is the best periodic sequence in the long run in the original Parrondo games. It can also be generalised to multi-player Parrondo games with strategy showing that the optimal strategy depends on the number of players.

The solution provides an example that the optimal {\em a priori} protocol can be found by looking at a related problem in which the protocol or strategy may take into account the state of an infinitely large ensemble of copies of the original system.

Furthermore, the algorithm is quite general and may be applied to other Markov decision problems, as shown in the section about the Primary Parrondo Paradox. In fact, this algorithm can also be applied to a discretization of a stochastic system continuous in time and provide an approximation for the optimal control protocol \cite{dinisepl}. The type of stochastic control problems relevant to stochastic thermodynamics \cite{Seifert} fit in this scheme.

\section*{Acknowledgements}
The author wishes to thank J.M.R. Parrondo for fruitful discussion and generous advice on the manuscript. The work was financially supported by grants BFM2001-0291 from Direcci\'on General de Ense\~{n}anza Superior and FIS2004-00271 (MCyT), MOSAICO (Consolider MEC), and PR27/05-13923 (Santander/Complutense).


\end{document}